# On correctness of the two-level model for description of active medium in quantum plasmonics

**Running head: 4-level model for active media description**

E D Chubchev[1], E S Andrianov[1], A A Pukhov[1,2,3], A P Vinogradov[1,2,3] and A A Lisyansky[4,5]

[1]Dukhov Research Institute of Automatics, 22 Sushchevskaya, Moscow 127055, Russia
[2]Moscow Institute of Physics and Technology, Moscow Region, Dolgoprudny, Russia
[3]Institute for Theoretical and Applied Electromagnetics, RAS, Moscow, Russia
[4]Department of Physics, Queens College of the City University of New York, Queens, NY 11367
[5]The Graduate Center of the City University of New York, New York, NY 10016

E-mail: lisyansky@qc.edu

**Abstract**
In order to simplify the theoretical description of spasers, a gain medium is commonly represented by a two-level system. A realistic model, however, should have four levels. By using the Lindblad equations we develop a description of such a system and show that depending on ratios of the Rabi frequency and the rate of relaxation of the polarization, a four-level system may be reduced to one of two effective two-level systems that reproduce the key properties of a four-level system.

## 1. Introduction

The field of plasmonics, in which the advantages of plasmonics and quantum electronics are combined, has been growing explosively in recent years [1-3]. However, ohmic losses in metals hinder numerous applications of plasmonic systems. Various approaches have been suggested to overcome this problem. Among them are the recent method of the plasmon injection [4] and a more common approach that utilizes such active media as quantum dots, dye molecules, and quantum wells [3, 5-9]. Using active media allows one to not only compensate losses but even to achieve an amplification of electric field [8]. It has also led to the quality improvement of metamaterials of optical range in which an absolute value of the negative refraction index was increased [9].

In metamaterials, particles of active media couple with plasmonic particles [5, 6, 9, 10] that results in forming a spaser - nanoresonator interacting with active media [11-13]. A spaser is a plasmonic analog of a laser, in which the plasmon oscillations play the role of the resonator mode, and feedback is implemented by the induced radiation of an inverted active medium back into the mode.

In spasers, the population inversion of an active medium is usually described by the Maxwell-Bloch equations for a two-level system (TLS) [11, 12, 14, 15]. However, active media are not two-level but more complex systems with many levels or even bands [16, 17]. The band structure of active media was considered in a number of papers [5-7, 18-20]. This approach, however, requires complicated computer calculations. At the same time, such laser characteristics as the generation threshold, frequency pulling, and spike operation are usually described by the simplest two-level model qualitatively correctly. Nonetheless, it is well known that such an approach cannot correctly treat some physical phenomena that are critical for spaser operation. For example, a TLS cannot be inverted with optical pumping [14]. To describe pumping in a TLS model, one needs to introduce phenomenological terms into the equations. A more adequate, but complex model for describing spasers models the gain medium as a three- or four-level system [19, 21]. This approach properly reflects the main features of the active medium, such as coherent pumping and the lasing transition. However, the resulting system of equations (these usually are the rate equations or the Maxwell-Bloch equations) can have nonphysical solutions, such as negative populations [22, 23].

Under certain conditions, the behavior of three- and four-level gain media is similar to TLS's [14]. Though modeling an active medium by a TLS greatly simplifies calculations, the results obtained require special analysis and justification.

The dynamics of a spaser with a TLS has been shown to exhibit Rabi oscillations [12, 15]. These oscillations seem impossible in a four-level gain medium because achieving the condition necessary for Rabi oscillations in a TLS requires matching the plasmon frequency of metal nanoparticle with the working frequency of the gain medium. The latter should be equal to the pumping frequency. With a four-level system (FLS) as a gain medium, these frequencies are different and plasmons do not have sufficient energy to excite the upper level of the FLS. This raises the question of whether Rabi oscillations are possible in spasers with realistic active media.

In this paper, we show that problems with the description of an FLS arise due to the use of uncontrollable approximations, such as a phenomenological description of relaxation processes. We demonstrate that a description of a spaser with a four-level active medium can be correctly reduced to a description with an effective TLS by using the Lindblad equations [24]. This avoids unphysical results such as negative populations [22, 23]. We show that in a spaser with a four-level active medium, Rabi oscillations may occur and provide the conditions for this to happen.

## 2. Equations for a spaser with a four-level gain medium

Consider a spaser consisting of a plasmon nanoparticle and a four-level molecule located nearby. A schematic of energy levels of the molecule is shown in Fig. 1. We assume that the transitions from level $|0\rangle$ to level $|3\rangle$ and from level $|2\rangle$ to level $|1\rangle$ can be described within the dipole approximation. In the case of dye molecules, the rates of the transitions are $\Gamma_{32} \sim \Gamma_{10} \sim 2\cdot10^{13}$ s$^{-1}$ and $\Gamma_{21} \sim 2\cdot10^{9}$ s$^{-1}$ [19, 25]. Thus, transitions from $|3\rangle$ to $|2\rangle$ and from $|1\rangle$ to $|0\rangle$ are faster than $|0\rangle \to |3\rangle$ and $|2\rangle \to |1\rangle$ transitions by four orders of magnitude, other transitions are forbidden. The pump is an external field with frequency $\omega_{30}$, which excites molecules into the upper state $|3\rangle$, the frequency $\omega_{21}$ of the transition $|2\rangle \to |1\rangle$ is the spaser transition. For R800 dye molecules, the wavelengths corresponding to the frequencies $\omega_{21}$ and $\omega_{30}$ are 750 nm and 300 nm [19].

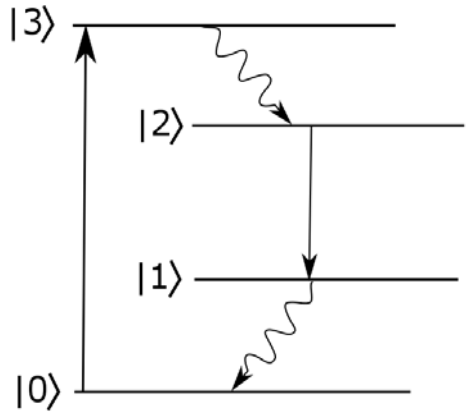

Fig.1. The schematic of the energy levels of the spaser gain medium.

The Hamiltonian of the plasmon mode of the metal nanoparticle has the form

$$\hat{H}_{SP} = \hbar\omega_{SP}\hat{a}^{\dagger}\hat{a}, \qquad (1)$$

where $\hat{a}^\dagger$ and $\hat{a}$ are Bose plasmon creation and annihilation operators, and $\omega_{SP}$ is the surface plasmon frequency, which is close to $\omega_{21}$. The Hamiltonian of the the molecule that we represent as a generic FLS has the form

$$\hat{H}_{mol} = \hbar\omega_{30}\hat{n}_3 + \hbar(\omega_{21}+\omega_{10})\hat{n}_2 + \hbar\omega_{10}\hat{n}_1. \tag{2}$$

where $\hat{n}_i = |i\rangle\langle i|$ is the operator of the population at the level $|i\rangle$ of the FLS. The total Hamiltonian of the system is

$$\hat{H} = \hat{H}_{SP} + \hat{H}_{mol} + \hat{V}. \tag{3}$$

The operator $\hat{V}$ describes the interaction of molecules with the nanoparticle and the pump field $\tilde{\mathbf{E}}_{pump}$. It has the form

$$\hat{V} = -\hat{\mathbf{d}}_M \cdot \left(\hat{\mathbf{E}}_{SP}(\mathbf{r}) + \tilde{\mathbf{E}}_{pump}\right), \tag{4}$$

where $\hat{\mathbf{E}}_{SP}$ is the operator of the electric field of the nanoparticle and $\mathbf{r}$ is the coordinate of the FLS. $\hat{\mathbf{E}}_{SP}$ and $\tilde{\mathbf{E}}_{pump}$ are defined by equations

$$\begin{aligned}\hat{\mathbf{E}}_{SP} &= \mathbf{E}(\mathbf{r})(\hat{a}^\dagger + \hat{a}), \\ \tilde{\mathbf{E}}_{pump} &= \mathbf{E}_{pump}(\exp[i\omega_{30}t] + \exp[-i\omega_{30}t]),\end{aligned} \tag{5}$$

where $\mathbf{E}(\mathbf{r})$ and $\mathbf{E}_{pump}$ are amplitudes of the plasmon mode and the pump field, respectively. Further, we use tilde to denote expectation values $\tilde{A}$ that can be represented as $\tilde{A} = A(\exp[i\omega t] + \exp[-i\omega t])$ or $\tilde{A} = A\exp[i\omega t]$, where $A$ is a slow varying amplitude and $\omega$ is a certain frequency.

In Eq. (4), the operator of the molecular dipole moment is defined as

$$\hat{\mathbf{d}} = \mathbf{d}_{21}\left(\hat{\sigma}_{21}^\dagger + \hat{\sigma}_{21}\right) + \mathbf{d}_{30}\left(\hat{\sigma}_{30}^\dagger + \hat{\sigma}_{30}\right), \tag{6}$$

where $\mathbf{d}_{21}$ and $\mathbf{d}_{30}$ are the matrix elements of the dipole moment for transitions $|2\rangle \to |1\rangle$ and $|0\rangle \to |3\rangle$, respectively, and the operators $\hat{\sigma}_{ij} = |i\rangle\langle j|$ are the transition operators from the state $|j\rangle$ to $|i\rangle$. Operators of the population $\hat{n}_j$ are associated with $\hat{\sigma}_{ij}$ through the relation $\hat{n}_j = \hat{\sigma}_{ij}^\dagger\hat{\sigma}_{ij} = \hat{\sigma}_{jj}$.

The dynamics of the system is described by the Lindblad equation [24]

$$\frac{d\hat{\rho}}{dt} = -\frac{i}{\hbar}\left[\hat{H},\hat{\rho}\right] + \hat{\hat{L}}\hat{\rho}, \tag{7}$$

where $\hat{\hat{L}}$ is a superoperator describing the relaxation of the FLS and the nanoparticle [26], and $\hat{\rho}$ is the density matrix of the whole "plasmon + active medium" system for which the Hilbert space is the direct product of Hilbert spaces of the plasmon and the FLS states. To preserve the positiveness of the density matrix, the operator $\hat{\hat{L}}\hat{\rho}$ must have the form [24]

$$\hat{\hat{L}}\hat{\rho} = \sum_{\substack{k,l=0\\k\neq l}}^{3}\frac{\Gamma_{kl}}{2}\left(2\hat{\sigma}_{kl}\hat{\rho}\hat{\sigma}_{kl}^\dagger - \hat{\rho}\hat{\sigma}_{kl}^\dagger\hat{\sigma}_{kl} - \hat{\sigma}_{kl}^\dagger\hat{\sigma}_{kl}\hat{\rho}\right) + \gamma_a\left(2\hat{a}\hat{\rho}\hat{a}^\dagger - \hat{a}^\dagger\hat{a}\hat{\rho} - \hat{\rho}\hat{a}^\dagger\hat{a}\right) \tag{8}$$

where $\gamma_a$ is the plasmon field relaxation rate, and $\Gamma_{kl}$ is the relaxation rate of the population for the transition from the state $|k\rangle$ to the state $|l\rangle$. The transition $|3\rangle \to |0\rangle$ competes with the transition $|3\rangle \to |2\rangle$. At the same time, $\Gamma_{30}$ is of the order of $10^9$ s$^{-1}$ which is much smaller than $\Gamma_{32}$ ($\sim 2\cdot 10^{13}$). Terms proportional to $\Gamma_{30}$ can, therefore, be neglected.

To obtain equations for expectation values, we multiply Eq. (7) by some operator $\hat{A}$ which does not explicitly depend on time. Then, we calculate a trace for the both part of the equation:

$$\frac{d}{dt}\text{tr}\left(\hat{A}\hat{\rho}\right) = -\frac{i}{\hbar}\text{tr}\left(\hat{A}\left[\hat{H},\hat{\rho}\right]\right) + \text{tr}\left(\hat{A}\hat{\hat{L}}\hat{\rho}\right) \tag{9}$$

To calculate $\text{tr}\left(\hat{A}[\hat{H},\hat{\rho}]\right)$ and $\text{tr}\left(\hat{A}\hat{\hat{L}}\hat{\rho}\right)$ it is covenient to use the trace invariance with respect to cyclic permutations:

$$\text{tr}\left(\hat{A}[\hat{H},\hat{\rho}]\right) = \text{tr}\left([\hat{A},\hat{H}]\hat{\rho}\right). \tag{10}$$

Analogously, we obtain

$$\text{tr}\left(\hat{A}\hat{\hat{L}}\hat{\rho}\right) = \sum_{\substack{k,l=0 \\ k \neq l}}^{3} \frac{\Gamma_{kl}}{2} \text{tr}\left\{\left(\hat{\sigma}_{kl}^{\dagger}[\hat{A},\hat{\sigma}_{kl}] - [\hat{A},\hat{\sigma}_{kl}^{\dagger}]\hat{\sigma}_{kl}\right)\hat{\rho}\right\} + \gamma_a \text{tr}\left\{\left(\hat{a}^{\dagger}[\hat{A},\hat{a}] - [\hat{A},\hat{a}^{\dagger}]\hat{a}\right)\hat{\rho}\right\}, \tag{11}$$

Replacing the operator $\hat{A}$ with the operators $\hat{\sigma}_{ij}$, $\hat{n}_i$, or $\hat{a}$ and using definitions of the respective operators and the commutation relations,

$$[\hat{\sigma}_{ij},\hat{\sigma}_{kl}] = \delta_{kj}\hat{\sigma}_{il} - \delta_{il}\hat{\sigma}_{kj}, \tag{12a}$$

$$[\hat{\sigma}_{ij},\hat{\sigma}_{kl}^{\dagger}] = \delta_{lj}\hat{\sigma}_{ik} - \delta_{ik}\hat{\sigma}_{lj}, \tag{12b}$$

$$[\hat{\sigma}_{ij},\hat{\sigma}_{ij}^{\dagger}] = \hat{n}_i - \hat{n}_j, \tag{12c}$$

$$[\hat{a},\hat{a}^{\dagger}] = 1, \tag{12d}$$

$$[\hat{a},\hat{\sigma}_{ij}] = [\hat{a},\hat{n}_i] = 0, \tag{12e}$$

we obtain equations for the respective averages.

To obtain relationships between the rates of longitudional and transverse relaxations, in Eq. (11), we substitute the operator $\hat{\sigma}_{ij}$ for $\hat{A}$ to obtain

$$\text{tr}\left(\hat{\sigma}_{ij}\hat{\hat{L}}\hat{\rho}\right) = \sum_{\substack{k,l=0 \\ k \neq l}}^{3} \frac{\Gamma_{kl}}{2} \text{tr}\left([\delta_{kj}\hat{\sigma}_{kl}^{\dagger}\hat{\sigma}_{il} - \delta_{il}\hat{\sigma}_{kl}^{\dagger}\hat{\sigma}_{kj} - \delta_{lj}\hat{\sigma}_{ik}\hat{\sigma}_{kl} + \delta_{ik}\hat{\sigma}_{lj}\hat{\sigma}_{kl}]\hat{\rho}\right)$$

$$= \sum_{\substack{k,l=0 \\ k \neq l}}^{3} \frac{\Gamma_{kl}}{2} \text{tr}\left([2\delta_{kj}\delta_{ki}\hat{\sigma}_{ll} - \delta_{il}\hat{\sigma}_{lj} - \delta_{lj}\hat{\sigma}_{il}]\hat{\rho}\right) = \sum_{l=0}^{3}\left(\delta_{ij}\Gamma_{il}n_l - \frac{\Gamma_{il}+\Gamma_{jl}}{2}\tilde{\sigma}_{ij}\right), \tag{13}$$

where $\tilde{\sigma}_{ij} = \text{tr}(\hat{\sigma}_{ij}\hat{\rho})$ and $n_i = \text{tr}(\hat{n}_i\hat{\rho})$. In equations for the polarization $\tilde{\sigma}_{ij}$ ($i \neq j$), the relaxation is described by the second term in the right-hand side of Eq. (15):

$$\text{tr}\left(\hat{\sigma}_{ij}\hat{\hat{L}}\hat{\rho}\right) = -\gamma_{ij}\tilde{\sigma}_{ij} = -\sum_{l=0}^{3}\frac{\Gamma_{jl}+\Gamma_{il}}{2}\tilde{\sigma}_{ij}, \tag{14}$$

where $\gamma_{ij} = \sum_{l=0}^{3}(\Gamma_{jl}+\Gamma_{il})/2$ is the decay rate of the polarization $\tilde{\sigma}_{ij}$. Eq. (14) connects the rates of relaxations of the polarization and the population. In particular, we have $\gamma_{21} = (\Gamma_{21}+\Gamma_{10})/2$ and $\gamma_{30} = \Gamma_{32}/2$.

By using Eq. (11) we can find the relaxation term corresponding to the population $n_i = \sigma_{ii}$:

$$\text{tr}\left(\hat{n}_i\hat{\hat{L}}\hat{\rho}\right) = \sum_{l=0}^{3}\Gamma_{li}n_l - \Gamma_{il}n_i. \tag{15}$$

Similarly, the relaxation term for the operator $\hat{a}$, Eq.(11), has the form

$$\text{tr}\left(\hat{a}\hat{\hat{L}}\hat{\rho}\right) = -\gamma_a\tilde{a}, \tag{16}$$

where $\tilde{a} = \text{tr}(\hat{\rho}\hat{a})$.

Using Eqs. (10) and (11) we obtain the equations for $\tilde{a}$, $\tilde{\sigma}_{21}$, $\tilde{\sigma}_{30}$, and $n_i$. We are interested in the regime with a large number of plasmons. In this regime, quantum fluctuations and correlations can be neglected. Therefore, averages of products of two operators can be split into products of averages of each of the operators. This is possible either in the case of developed spasing or if a large initial number of plasmons is excited by pulse pumping [27].

Then, to simplify our equations, we use the rotating wave approximation. This approximation can be used when the molecular transition frequency $\omega_{21}$ is close to the plasmon resonance frequency $\omega_{SP}$. In the framework of this approximation, we can express $\tilde{a}$, $\tilde{\sigma}_{21}$, and $\tilde{\sigma}_{30}$ through slow amplitudes $\tilde{a}$, $\sigma_{21}$, and $\sigma_{30}$, as $\tilde{a} = a\exp[i\omega t]$, $\tilde{\sigma}_{21} = \sigma_{21}\exp[i\omega_{21}t]$ and $\tilde{\sigma}_{30} = \sigma_{30}\exp[i\omega_{30}t]$, where $\omega$ is the generation frequency. Neglecting the fast-oscillating terms with frequencies $\pm 2\omega_{21}$, $\pm 2\omega_{30}$, $\pm(\omega_{30}+\omega)$, we obtain the Maxwell-Bloch equations for the FLS for $c$-numbers

$$\dot{a} = -i\Omega_{21}\sigma_{21} + (i\delta_1 - \gamma_a)a, \tag{17a}$$

$$\dot{\sigma}_{21} = i\Omega_{21}aD + (i\delta - \gamma_{21})\sigma_{21}, \tag{17b}$$

$$\dot{D} = 2i\Omega_{21}(a^*\sigma_{21} - a\sigma_{21}^*) - 2\Gamma_{21}(D+n_1) + \Gamma_{32}n_3 + \Gamma_{10}n_1, \tag{17c}$$

$$\dot{n}_1 = -i\Omega_{21}(a^*\sigma_{21} - a\sigma_{21}^*) + \Gamma_{21}(D+n_1) - \Gamma_{10}n_1, \tag{17d}$$

$$\dot{n}_3 = i\Omega_{30}(\sigma_{30} - \sigma_{30}^*) - \Gamma_{32}n_3, \tag{17e}$$

$$\dot{\sigma}_{30} = i\Omega_{30}(n_3 - n_0) - \gamma_{30}\sigma_{30}, \tag{17f}$$

$$n_0 + n_1 + (D + n_1) + n_3 = 1, \tag{17g}$$

where $\Omega_{21} = \mathbf{d}_{21}\cdot\mathbf{E}(\mathbf{r})/\hbar$ and $\Omega_{30} = \mathbf{d}_{30}\cdot\mathbf{E}_{pump}/\hbar$ are the coupling constants of the FLS with the fields of the nanoparticle and pumping, respectively, $D = n_2 - n_1$ is the population inversion, variables $\delta_1 = \omega - \omega_{SP}$ and $\delta = \omega - \omega_{21}$ are the detunings of the lasing frequency $\omega$ from the frequency of the plasmon mode $\omega_{SP}$ and the frequency of the working transition $\omega_{21}$, respectively. Below, we consider small detuning so that both $\delta_1$ and $\delta$ are much smaller than $\gamma_{21}$. In Eqs. (17), the condition for the normalization of the density matrix, $n_0 + n_1 + n_2 + n_3 = 1$, is taken into account.

## 3. Reduction of the four-level system to a two-level

In the case when transitions $|1\rangle \to |0\rangle$ and $|3\rangle \to |2\rangle$ occur faster than all other processes in a molecule, the system of equations for the four-level gain medium (17) can be reduced to an effective TLS [14, 15]

$$\dot{a} = -i\Omega_{21}\sigma_{21} + (i\delta_1 - \gamma_a)a, \tag{18a}$$

$$\dot{\sigma}_{21} = i\Omega_{21}aD + (i\delta - \gamma_{21})\sigma_{21}, \tag{18b}$$

$$\dot{D} = 2i\Omega_{21}(a^*\sigma_{21} - a\sigma_{21}^*) - (\Gamma_{21} + W'_{pump})D + (W'_{pump} - \Gamma_{21}), \tag{18c}$$

where $W'_{pump}$ descibes incoherent pumping. Though such an approach is widely used, it is based on a phenomenological description of relaxation, which may lead to unpredictable results [22, 23]. We rederive the TLS approach by using on the Lindblad equation (9).

To simplify Eqs. (17), we consider the case of a weak pump field which satisfies the inequality $\Omega_{30} \ll \Gamma_{32}$. This regime is usually realized in experiments for most of lasers. The intensity of pumping at which $\Omega_{30} \sim \Gamma_{32}$ is about a 4 MW/cm$^2$. At such a pump rate, at the transition $|0\rangle \to |3\rangle$, the Rabi oscillations may occur and, therefore, an FLS cannot be reduced to a TLS. Therefore, we only consider the regime $\Omega_{30} \ll \Gamma_{32}$.

If $\Omega_{30} \ll \Gamma_{32}$, then $\sigma_{30}$ and $n_3$ adiabatically adjust to the instantaneous values of other variables. Therefore, their time derivatives can be neglected:

$$\sigma_{30} = \frac{i\Omega_{30}}{\gamma_{30}}(n_3 - n_0), \tag{19a}$$

$$n_0 = \left(1 + \frac{\gamma_{30}\Gamma_{32}}{2\Omega_{30}^2}\right)n_3. \tag{19b}$$

The condition $\Omega_{30} \ll \Gamma_{32}$ is necessary for the reduction of an FLS to an effective TLS. If the condition $\Omega_{30} \ll \Gamma_{32}$ is not fulfilled, $n_3$ and $\sigma_{30}$ cannot be excluded from Eqs. (17). Thus, this condition is necessary for the reduction of an FLS to an effective TLS. As we show below, depending on the relationship between the coupling constant of the field with the working transition, $\Omega_{21}|a|$, and the rate of depletion of the lower level of this transition, $\Gamma_{10}$, we need to consider separately the cases of strong, $2\Omega_{21}|a| \geq \Gamma_{10}$, and weak, $2\Omega_{21}|a| \ll \Gamma_{10}$, couplings.

For the weak coupling, the population $n_1$ adjusts to the instantaneous values of the other variables in the same way as $\sigma_{30}$ and $n_3$:

$$n_1 = \frac{-i\Omega_{21}(a^*\sigma_{21} - a\sigma_{21}^*) + \Gamma_{21}D}{\Gamma_{10} - \Gamma_{21}} \tag{19c}$$

At the wavelength of 650 nm for a silver nanosphere, $\Omega_{21} < 10^{11}$ s$^{-1}$ [15], which is smaller than $\Gamma_{10}$. Substituting Eqs. (18a) and (18c) into Eq. (17c), one obtains the equation for the population inversion of the effective TLS:

$$\dot{D} = \alpha i\Omega_{21}(a^*\sigma_{21} - a\sigma_{21}^*) - \beta\Gamma_{21}(D - D_0), \tag{20}$$

where

$$\alpha = 1 + \frac{\Gamma_{21}}{\Gamma_{10} - \Gamma_{21}} + \frac{\Gamma_{32}}{\Gamma_{10} - \Gamma_{21}} \frac{1}{1 + \gamma_{30}\Gamma_{32}/(2\Omega_{30})^2}, \tag{21a}$$

$$\beta = 1 + \frac{\Gamma_{21}}{\Gamma_{10} - \Gamma_{21}} + \frac{\Gamma_{32}}{\Gamma_{21}} \frac{\Gamma_{10} + \Gamma_{21}}{\Gamma_{10} - \Gamma_{21}} \frac{1}{1 + \gamma_{30}\Gamma_{32}/(2\Omega_{30})^2}, \tag{21b}$$

and $D_0$ is the steady state population inversion which is equal to

$$D_0 = \frac{1}{2\beta\Gamma_{21}} \frac{\Gamma_{32}}{1 + \gamma_{30}\Gamma_{32}/(2\Omega_{30})^2}. \tag{21c}$$

Because $\Gamma_{21} \ll \Gamma_{10}$, $\Omega_{30} \ll \Gamma_{32}$, and $\Omega_{30} \sim \Gamma_{21}$, the quantities $\alpha$ and $\beta$ are close to unity. Note, that in contrast to the standard model of a TLS (17), $D_0$ is always non-negative [see Eq. (20c) and Ref. 12]. Also, in the weak-coupling-regime, the coefficient $\alpha \approx 1$ is two times smaller than the value given by the James-Cummings model for standard single-mode equations [14].

After adiabatic elimination of $n_1$, $n_3$, and $\sigma_{30}$ [see Eq. (19a)-(19c)], only three equations remain:

$$\dot{a} = -i\Omega_{21}\sigma_{21} + (i\delta_1 - \gamma_a)a, \tag{22a}$$

$$\dot{\sigma}_{21} = i\Omega_{21}aD + (i\delta - \gamma_{21})\sigma_{21}, \tag{22b}$$

$$\dot{D} = \alpha i\Omega_{21}(a^*\sigma_{21} - a\sigma_{21}^*) - \beta\Gamma_{21}(D - D_0). \tag{22c}$$

In the weak coupling limit, Eqs. (22) can be simplified further. Since we consider the case $2\Omega_{21}|a| \ll \Gamma_{10}$ and $\gamma_{21} = (\Gamma_{10} + \Gamma_{21})/2$, the relaxation process is more rapid than oscillations in Eq. (22b) ($\gamma_{21} \gg \Omega_{21}|a|$). Therefore, similar to time derivatives of $n_1$, $n_3$, and $\sigma_{30}$, the time derivative of the polarization $\sigma_{21}$ can be neglected to give

$$\sigma_{21} = \frac{i\Omega_{21}D}{\gamma_{21} - i\delta}a. \tag{23}$$

Substituting Eqs. (23) into Eqs. (22a) and (22c), one obtains the following system of two equations describing the dynamics of a spaser with an effective TLS:

$$\dot{D} = -2\alpha\frac{\Omega_{21}^2\gamma_{21}}{\gamma_{21}^2 + \delta^2}D|a|^2 - \beta\Gamma_{21}(D - D_0), \tag{24a}$$

$$\dot{a} = \frac{\Omega_{21}^2 Da}{\gamma_{21} - i\delta} + (i\delta_1 - \gamma_a)a. \tag{24b}$$

The weak coupling, $2\Omega_{21}|a| \ll \Gamma_{10}$, imposes restrictions on the distance between the dye molecule and the nanoparticle. The critical value of the distance $r_{cr}$ at which the transition from the weak to the strong coupling limit occurs is determined by the formula $\Omega_{21}|a| = \gamma_{21}$. By the definition of the coupling constant, $\Omega_{21}|a| = \mathbf{d} \cdot \mathbf{E}/\hbar$, where $\mathbf{E} \sim \mathbf{d}_{NP}/r^3$, $\mathbf{d}_{NP}$ is the nanoparticle dipole moment which is proportional to $|a|$, $r$ is the distance between the nanoparticle and the dye molecule. Using the definition of the coupling constant, we can obtain the equation for $r_{cr}$

$$r_{cr} = \left( \frac{d_{NP} d}{\hbar \gamma_{21}} \right)^{1/3}. \tag{25}$$

A dipole moment of a dye molecule varies in the range from 10 D to 100 D, a dipole moment of the nanoparticle is about 200 D, and the polarization decay rate $\gamma_{21}$ is of the order of $10^{13} - 10^{14}$ c$^{-1}$ [19, 28-33]. Therefore, $r_{cr}$ is about 4 or 20 nm. The electric field strength at such distances is of the order of $10^7 - 10^8$ V/m. Though these values are too small for the dielectric breakdown in the CW regime, at such values of the intensity a destruction of a gain medium is possible. However, if pumping is done by a pulse with small duration and intensity, such a destruction can be prevented.

Now, let us consider a strong coupling, $2\Omega_{21}|a| \gg \gamma_{21}$. In this case, the time derivatives of $\sigma_{21}$ and $n_1$ in Eqs. (17) cannot be neglected. However, we can equate to zero time derivatives of $n_3$ and $\sigma_{30}$ assuming that their deviations from stationary values are negligible at any time. This approximation is correct if the populations of the upper levels of a molecule are small, $n_1, n_2, n_3 \ll 1$. As a consequence, substituting $n_0$ from Eq. (17g) into Eq. (17f) and $\sigma_{30}$ into Eq. (17e) we obtain

$$-i\Omega_{30} - \frac{\Gamma_{32}}{2} \sigma_{30} \approx 0, \tag{26}$$

$$\dot{n}_3 = \frac{4\Omega_{30}^2}{\Gamma_{32}} - \Gamma_{32} n_3. \tag{27}$$

Keeping in mind that the deviation of $n_3$ from the steady value is zero, one obtains, that $n_3^{st} = (2\Omega_{30}/\Gamma_{32})^2$. Finally, we arrive at

$$\dot{a} = -i\Omega_{21}\sigma_{21} + (i\delta_1 - \gamma_a)a, \tag{28a}$$

$$\dot{\sigma}_{21} = i\Omega_{21} a D + (i\delta - \gamma_{21})\sigma_{21}, \tag{28b}$$

$$\dot{D} = 2i\Omega_{21}(a^*\sigma_{21} - a\sigma_{21}^*) - (\Gamma_{10}/2 + \Gamma_{21})D + (\Gamma_{10}/2 - \Gamma_{21})n + W_{pump}, \tag{28c}$$

where $W_{pump} = 4\Omega_{30}^2/\Gamma_{32}$ is the effective pump rate for the level $|2\rangle$ and $n = n_2 + n_1 = 2n_1 + D$ is the total population at the working transition, which is described by the equation

$$\dot{n} = W_{pump} - \Gamma_{10}(n - D)/2. \tag{28d}$$

Thus, in the strong coupling limit, we reduce spaser equations for a four-level active medium to an active medium modeled by an effective TLS, Eqs. (28).

Since the normalization condition $n_2 + n_1 = 1$ for the effective TLS is not satisfied because $n_2 + n_1 = 1 - n_0 - n_3$, the following differences between systems of Eqs. (17) and (28) arise: (*i*) the latter system contains the additional equation for $n = n_1 + n_2$ [Eq. (28d)], (*ii*) in Eq. (28c), a new term $(\Gamma_{10}/2 - \Gamma_{21})n$ appears due to the relaxation process from the level $|1\rangle$ to the level $|0\rangle$, and (*iii*) $D_0$ is non-negative because of slow relaxation process from $|2\rangle$ to $|1\rangle$. In the next section. we consider the changes in a spaser with a four-level gain medium due to these differences.

## 4. Comparison of the dynamic of a TLS and an effective TLS in spasers
In Ref. 13, oscillations of the number of plasmons in a TLS-spaser during the transition to a steady state were predicted. These oscillations were identified as Rabi oscillations caused by the exchange of quanta between the plasmon mode and the gain medium [14]. Existence of Rabi oscillations in a TLS is well known [15], but in a spaser with a four-level gain medium whose operating frequency coincides with the transition frequency between two excited states, such an exchange of quanta seems impossible due to the rapid relaxation of the lower level of the spaser transition.

The Rabi oscillations in a spaser should be distinguished from the relaxation oscillations. The relaxation oscillations may be observed at the initial stages of lasing, Their existence requires CW pumping, whereas the Rabi oscillations in spaser may be observed at initial pulse pumping. The frequency of relaxation oscillations is proportional to the geometric mean between the decay rate of the field and the longitudinal decay rate [14]. Thus, the relaxation oscillations does not directly connected to the level structure of the system and could be observed both in two-level and four-level schemes. In contrast to the relaxation oscillations, the Rabi oscillations arise at subthreshold pump intensities [15]. In spaser the Rabi oscillations arise due to nonradiative resonance energy interchange by excitations between plasmonic particle and atom [20].

In the case of the weak coupling, $\Omega_R = 2\Omega_{21}|a| \ll \Gamma_{10}$, a spaser is described by Eqs. (24). For such a value of the Rabi frequency, the molecular polarization simply adjusts to the plasmon field [see Eq. (23)] and, therefore, Rabi oscillations do not occur. To demonstrate the absence of the Rabi oscillations, we assume that the surface plasmon amplitude $a$ is a constant and consider the behavior of an effective TLS in a given field. Then, the dynamics of the population inversion is described by Eq. (24a). The solution of this equation has the form

$$D(t) = \left(D(0) - D_{st}\right)\exp\left[-\left(\alpha\frac{\Omega_{21}^2 \gamma_{21}}{\gamma_{21}^2 + \delta^2}|a|^2 + \beta\Gamma_{21}\right)t\right] + D_{st} \qquad (29)$$

where

$$D_{st} = \beta\Gamma_{21}\left[\alpha\Omega_{21}^2 \gamma_{21}\left(\gamma_{21}^2 + \delta^2\right)^{-1}|a|^2 + \beta\Gamma_{21}\right]^{-1} D_0$$

is the stationary value of the population inversion in an external field. Eq. (29) describes a monotonic transition to a steady state in a given field.

If the coupling is strong, $2\Omega_{21}|a| \gg \Gamma_{10}$, a spaser with a four-level gain medium is described by Eqs. (28). To consider Rabi oscillations for the system of equations for the effective TLS, we assume for simplicity that in Eqs. (18a,b) and (29a,b), $\delta = \delta_1 = 0$. Then, the value of the polarization $\sigma_{21}$ is purely imaginary provided that the strength of the field $a$ is real. At the initial moment, $D$ has a steady value corresponding to the intensity of the pump in the absence of the field at the frequency $\omega_{21}$. The pump rate is assumed to be lower than that required for lasing. If the pump intensity is zero, then the gain medium cannot interact with the plasmon mode. Fig. 2 obtained as a numerical solution of the system of equations (18) and (28) shows the dynamics of the transition to a steady state of spasers with four-level and two-level gain media. The initial value $a(0)$ is chosen such that $2\Omega_{21}|a(0)| \gg \Gamma_{10}$. As shown in Fig. 2, the population inversion $D$ of the four-level spaser oscillates. The oscillations are damped in the time interval of the order of $\Gamma_{10}^{-1}$. Rabi oscillations of the four-level spaser are damped slightly faster than that of the TLS spaser because the total population of the spaser transition levels, $n_1 + n_2$, decreases due to the transition from level $|1\rangle$ to $|0\rangle$. Rabi oscillations occur because their period is larger than the rate of the transition $|1\rangle \to |0\rangle$. This is possible due to the high intensity of the field acting on the gain medium. In other words, the first level cannot clear up completely during the period of Rabi oscillations.

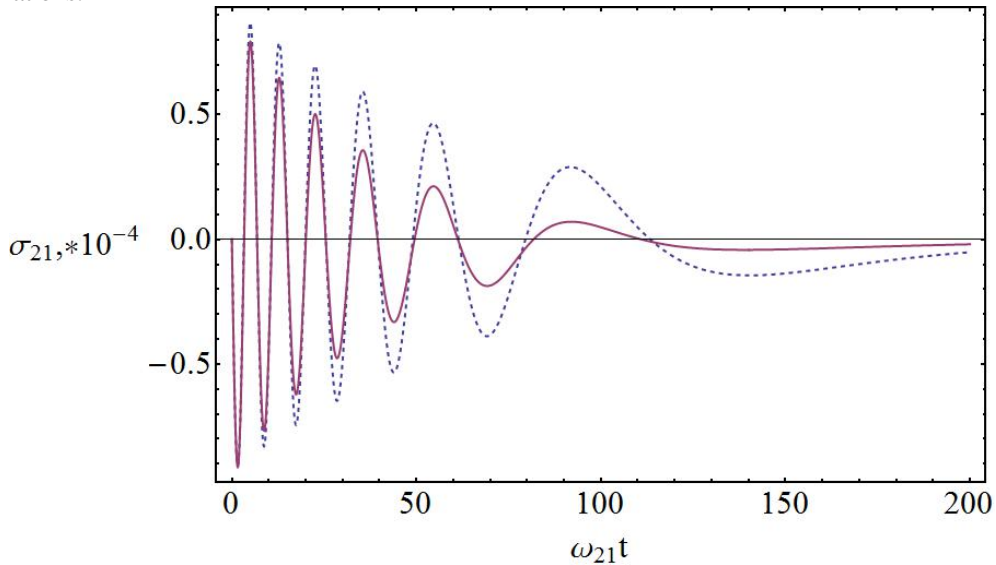

Fig. 2. Polarizations of a TLS spaser (the dashed line) and an effective TLS spaser (the solid line) vs time. The curve for the FLS coincides with the curve of the effective TLS spaser. The initial conditions for the four-level spaser are $a(0) = 20i$, $n(0) = n_0$, $D(0) = D_0$, and $\sigma_{21}(0) = 0$, where $n_0 = W_{pump}\left(1/\Gamma_{21} + 1/\Gamma_{10}\right)$ and $D_0 = W_{pump}\left(1/\Gamma_{21} - 1/\Gamma_{10}\right)$ are the total population and the population inversion in the steady state. The initial conditions for the two-level spaser are $a(0) = 20i, D(0) = D_0$, and $\sigma_{21}(0) = 0$. The pump intensity corresponds to the steady-state inversion value $D_0 = 1.8 \cdot 10^{-4}$. This value of $D_0$ can be achieved at the pump power 150 mW/cm$^2$.

Thus, unlike to the conventional TLS, in a spaser built on a four-level active medium, to observe the Rabi oscillations, an arbitrarily small pump power is required. Thereby, the phenomenon of Rabi oscillations predicted within the framework of the TLS model is not an artifact but it is preserved in a more detailed description of an active medium.

## 5. Conclusion

We have described a realistic spaser with a four-level gain medium by using the Lindblad equations. This approach assures that the populations of gain medium levels are always positive. In the framework of this approach, the four-level gain medium can be reduced to an effective two-level medium for experimentally feasible pump rates. We show that the effective TLS is described by two different systems of equations which depend on the ratio of the Rabi frequency and the decay rate of the polarization. We also show that a spaser composed of a plasmonic nanoparticle and the effective TLS gain medium behaves in fashion similar to that of a spaser composed of the same plasmonic nanoparticle and a four-level gain medium. In particular, in the strong coupling regime, the Rabi oscillations predicted by the TLS model are possible in an FLS as well.

**Acknowledgement**
A.A.L. would like to acknowledge support from the NSF under Grant No. DMR-1312707.